\documentclass[12pt,english]{article}

\usepackage[latin9]{inputenc}
\usepackage{geometry}
\usepackage{prettyref}
\usepackage{amsmath}
\usepackage{amssymb}
\usepackage{graphicx}
\usepackage{esint}

\usepackage{caption}

\usepackage{slashed}

\usepackage{dsfont}

\usepackage{color}
\definecolor{darkgreen}{rgb}{0,0.5,0}
\definecolor{darkblue}{rgb}{0,0,0.6}
\definecolor{purple}{rgb}{0.4,.2,0.7}
\usepackage[colorlinks=true,citecolor=darkgreen,linkcolor=purple,urlcolor=purple]{hyperref}

\numberwithin{equation}{section}
\numberwithin{figure}{section}
\numberwithin{table}{section}

\def\CN{{\cal N}}

\DeclareFontShape{OT1}{cmr}{mx}{n}{<->cmr10}{}

\begin{document}

\fontseries{mx}\selectfont

\newcommand{\f}[1]{\textcolor{blue}{#1}}
\newcommand{\fc}[1]{\textcolor{red}{#1}}
\newcommand{\zimo}[1]{\textcolor{purple}{#1}}

\centerline{\Large{Higher genus Siegel forms and multi-center black holes }}
\smallskip\smallskip
\centerline{\Large{in ${\cal N}=4$ supersymmetric string theory}}
\bigskip
\bigskip
\centerline{Frederik Denef,$^1$ Shamit Kachru,$^2$ Zimo Sun,$^1$ and Arnav Tripathy$^3$}
\bigskip
\bigskip
\centerline{$^1$ Department of Physics, Columbia University}
\centerline{538 West 120th Street, New York, NY 10027}
\medskip
\centerline{$^2$ Stanford Institute for Theoretical Physics}
\centerline{Stanford University, Palo Alto, CA 94305}
\medskip
\centerline{$^3$ Department of Mathematics, Harvard University}
\centerline{Cambridge, MA 20138}
\bigskip
\bigskip

\noindent
{\bf Abstract:} 
We conjecture that the Fourier coefficients of a degree three Siegel form, $1/\sqrt{\chi_{18}}$, count the degeneracy
of three-center BPS bound states in type II string theory compactified on $K3 \times T^2$.  We provide
evidence for our conjecture in the form of consistency with physical considerations of wall-crossing, holographic bounds, and the appearance
of suitable counting functions (involving the inverse of the modular discriminant $\Delta$ and the inverse of the Igusa cusp form $\Phi_{10}$) 
in limits where the count degenerates to involve single-center or two-center objects. 

\newpage

\tableofcontents

\section{Introduction}

The microscopic accounting of black hole entropy for suitably simple (BPS) black holes \cite{StromingerVafa} is one of the triumphs of string theory.  Hints of deep
ties between this subject and natural objects in number theory and the theory of automorphic forms began to appear with the work of Dijkgraaf, Verlinde and
Verlinde \cite{DVV}, where a degree two Siegel modular form -- the Igusa cusp form
$\Phi_{10}$ -- appears and plays a central role in determining the microstate counts
in ${\cal N}=4$ supersymmetric string theory on $K3 \times T^2$.
A heuristic explanation of the appearance of a hidden genus two curve in this problem
was provided by Gaiotto \cite{Gaiotto}, and important further developments are
reviewed in, for instance, \cite{Sen, DMZ}.

In a parallel line of development, it was soon realized that a given charge sector
in supersymmetric string theory may also support multi-center BPS black hole configurations \cite{Denef:2000nb, Denef:2002ru, Bates:2003vx}.  These play a crucial role in resolving various
paradoxes with the attractor mechanism for BPS black holes \cite{FKS,mooreattractor}, and appear in the 
proper interpretation of the states the Igusa cusp form (or more properly $1/\Phi_{10}$) is enumerating \cite{Sen:2007vb,Dabholkar:2007vk,Sen:2007pg,Cheng:2007ch,Banerjee:2008yu,Sen:2008ht,Dabholkar:2009dq}.    

In this paper, we further tie these lines of development together by proposing that
there is a preferred
degree three Siegel form whose Fourier coefficients are counting the microstates of
three-center BPS solutions in type II string theory on $K3 \times T^2$. 
It will not escape the attention of the reader that to the extent our conjecture holds, it suggests a family of conjectures capturing higher multi-center degeneracies
as well.

The organization of our exposition is as follows.  In section \ref{sec:Siegel}, we introduce the hero of our story, the degree three
Siegel form $\chi_{18}$. In section \ref{sec:conjecture-rough} we give a first formulation of our conjecture tying the Fourier coefficients of $1/\sqrt{\chi_{18}}$ to degeneracies of 
BPS states, ignoring subtleties related to wall crossing. In section \ref{sec:threeparticleboundstates}, we make some of the ingredients introduced here more precise, and
review some relevant properties of three-center BPS bound states, including BPS degeneracies of three-node quivers corresponding to the Higgs branch of three-center ``scaling solutions'' found in earlier work \cite{Denef:2007vg,Bena:2012hf,Lee:2012sc,Manschot:2013sya}.
In section \ref{sec:conjecture}, we formulate a more precise version of the conjecture, taking into account wall crossing ambiguities. Sections \ref{sec:leading-q-examples} and \ref{sec:higher-order} witness various tests of the conjecture -- with tests of wall-crossing appearing in
\ref{sec:leading-q-examples}, and limits yielding $1/\Delta$ and $1/\Phi_{10}$ tested in section \ref{sec:higher-order}.  We close with a discussion of (admittedly
big) open
questions in section \ref{sec:discussion}.

\section{An interesting degree three Siegel form} \label{sec:Siegel}

We begin with the observation that the known counting functions for BPS state degeneracies on $K3 \times T^2$ have simple relations to
bosonic string partition functions.  Recall that the counting function for $1/2$-BPS Dabholkar-Harvey states is given by
$$Z_{\rm genus \,\, one} \equiv {1\over \Delta} = q^{-1} \prod_n {1 \over (1-q^n)^{24}}$$
where we have indicated that the inverse of $\Delta$ is also the (chiral) 1-loop bosonic partition function.
Similarly,
$$Z_{\rm genus\,\, two} \equiv {1\over \Phi_{10}}$$
also arises as the genus two chiral measure in the bosonic string.  
So suggestively, the genus one partition function counts 1/2-BPS objects, realized in supergravity as single-center black holes, while the genus two partition function counts bound states of two such $1/2$-BPS objects, realized in supergravity as black hole configurations with up to two centers.

\medskip
These facts motivate, as a natural guess, the analogous genus-three measure
$$ Z_{\rm genus\,\, three} \equiv \frac{1}{\sqrt{\chi_{18}}} \equiv \frac{1}{\chi_9}$$ as the counting function for bound states of three 1/2-BPS objects, realized in supergravity as black hole configurations with up to three centers.
This function was found to occur in the genus three partition function in \cite{BKMP}.  It is easiest to describe as a product of theta functions with
characteristic, in the following way.

\medskip
On a compact Riemann surface $X$ of genus $g$, one can choose a symplectic basis for $H_1(X,{\mathbb Z})$ $a_1, \cdots a_g, b_1, \cdots, b_g$.
Then a basis of holomorphic 1-differentials is determined by the conditions
$$\int_{a_i} \omega_j = \delta_{ij}~.$$
The period matrix of the surface is then fixed by
$$\int_{b_i} \omega_j = \tau_{ij},$$
with $\tau$ symmetric and ${\rm Im}(\tau) > 0.$  $\tau$ gives the parametrization of Riemann surfaces $X$ by the Siegel upper half space.

\medskip
In terms of this data,
recall that the genus $g$ theta functions are given by (see e.g.\ \cite{Beilinson})
$$\theta(z,\tau) = \sum_{m \in {\mathbb Z}^g} {\rm exp}\left( 2\pi i (m^t z + {1\over 2}m^t \tau m)\right)~.$$
Here, $\theta: {\mathbb C}^g \times H_g \to {\mathbb C}$ takes as arguments $z \in {\mathbb C}^g$ and the $g \times g$ period matrix
$\tau$.

\medskip
The theta functions with characteristic can be similarly defined.  Let
$$\alpha = \epsilon + \tau \delta \in {\mathbb C}^g$$
with $\epsilon, \delta$ taking values in ${1\over 2}{\mathbb Z}^g$.  With the Jacobian of $X$ being $J = {\mathbb C}^g / T$,
the class of $\alpha$ in $J$ is called a theta characteristic.
We define the parity of $\alpha$ as $4 \epsilon \delta ~{\rm mod} ~2$.
In particular, of the $4^g$ possible choices of characteristic, we then have $2^{g-1} (2^g - 1)$ odd ones and $2^{g-1} (2^g + 1)$ even ones.
Now, define
$$\theta[\alpha](z,\tau) = {\rm exp}\left(2\pi i(\delta^t (z + \epsilon) + {1\over 2} \delta^t \tau\delta) \right) \theta(z+\epsilon + \tau\delta,\tau)~.$$
These are the desired theta functions with characteristic. We will only need these evaluated at $z=0$, and denote $\theta[\alpha] \equiv \theta[\alpha](\tau) \equiv \theta[\alpha](z,\tau)|_{z=0}$.

\medskip
The automorphic forms $\Delta$, $\Phi_{10}$ and $\chi_{9}=\sqrt{\chi_{18}}$ all have simple definitions as products of genus $g=1,2,3$ theta functions with characteristics:
\begin{align}
 \Delta &= 2^{-8} \prod_{\alpha \,\,{\rm even}} \theta[\alpha]^8 \qquad \,\,\, (g=1) \\
 \Phi_{10} &= 2^{-12} \prod_{\alpha \,\,{\rm even}} \theta[\alpha]^2 \qquad \, (g=2) \\
 \chi_{9} &= 2^{-14} \prod_{\alpha \,\,{\rm even}} \theta[\alpha]^{\frac{1}{2}}  \qquad (g=3) 
\end{align}
As there are 36 even theta functions with characteristic at genus three, $\chi_{18}$ indeed defines an automorphic form of weight 18.


Our claim --- which we make more precise below --- is that $Z_{g=3} = 1/\chi_9$ gives the BPS counting function for bound states of three 1/2-BPS constituents on $K3 \times T^2$, corresponding to black hole configurations with up to three centers.

\section{Conjecture} \label{sec:conjecture-rough}

In this section we give a first formulation of our conjecture, without being precise about the moduli-dependence of bound state degeneracies, and without being precise about various sign ambiguities. We will likewise deliberately be vague about the distinction between BPS indices and absolute degeneracies. In section \ref{sec:conjecture} we will give a more precise version of the conjecture, and argue that the sign ambiguities and moduli-dependence are in fact closely related. 

\begin{figure}
\begin{center}
 \includegraphics[width=\textwidth]{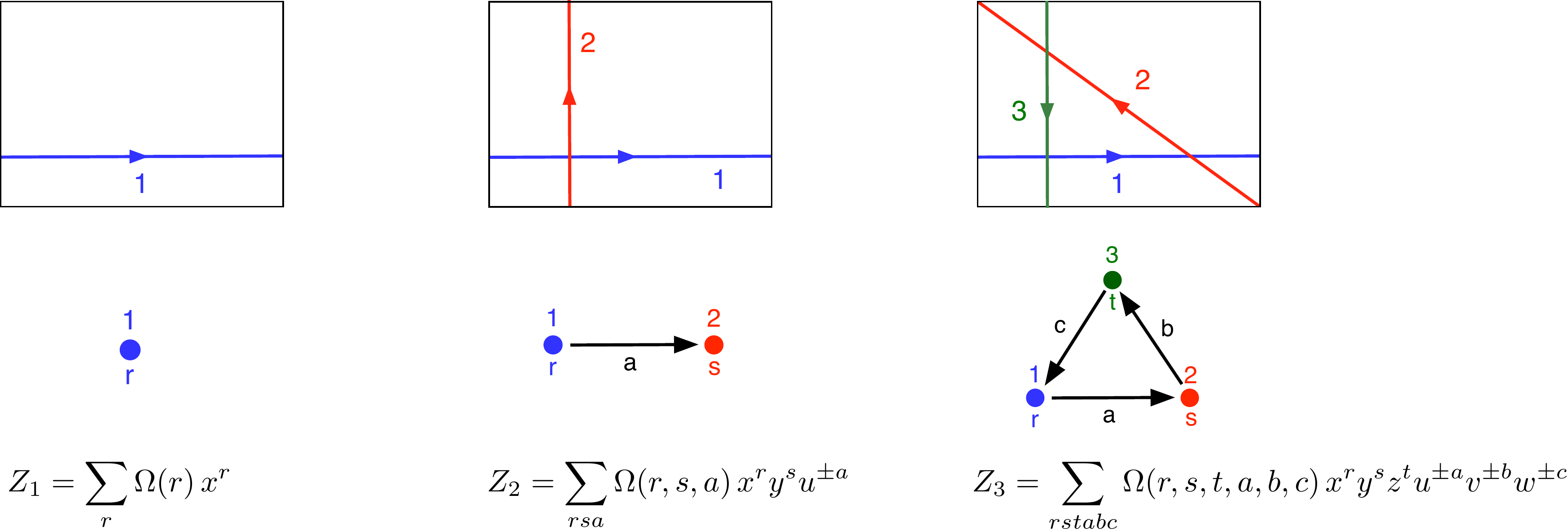}
\end{center}
\caption{D-brane configurations, quiver diagrams encoding intersection numbers, and bound state counting functions for $g=1$, $g=2$ and $g=3$. \label{g123}}
\label{branes}
\end{figure}

We first review the interpretation of $Z_{g=1}$ and $Z_{g=2}$ as BPS bound state counting functions and then state our proposed generalization for $Z_{g=3}$. The basic idea and our notations are illustrated in fig.\ \ref{g123}. For concreteness we consider type IIB string theory compactified on $T^2 \times K3$. In a suitable duality frame, 1/2-BPS states are represented as D-branes wrapping a 1-cycle $\gamma$ on $T^2$ and an even-dimensional cycle $Q \in H_{\rm even}(K3) = H_0(K3) \oplus H_2(K3) \oplus H_4(K3)$. The number of BPS states $\Omega$ with charge $\Gamma = \gamma \times Q$ depends only on the duality invariant 
\begin{align}
 r \equiv \frac{1}{2} Q \cdot Q \, ,
\end{align}
where the dot product denotes the signature $(4,20)$ intersection product on $H_{\rm even}(K3)$. The generating function for the degeneracies $\Omega(r)$ is $Z_{g=1} = 1/\Delta$, expanded in powers of
\begin{align}
 x \equiv e^{2 \pi i \tau} \, ,
\end{align}
that is to say,
\begin{align} \label{Zg1}
 Z_{g=1}(x) = \frac{1}{\Delta(x)} = \sum_r \Omega(r) \, x^r = x^{-1} + 24 \, x^0 + 324 \, x + 3200 \, x^2 + \cdots 
\end{align}   
In the 4d $\CN=4$ low energy effective supergravity theory, these 1/2-BPS states are realized as ``small'' single-center black holes. The minimal value of $r=\frac{1}{2}Q^2$ is $-1$. This corresponds to a rigid cycle $Q$ wrapped in K3, for example K3 itself or a supersymmetric 2-sphere. Higher $r$ correspond to wrapped branes with deformation moduli. A smooth genus $g$ supersymmetric 2-cycle has $r=g-1$. 

Next we consider two 1/2-BPS branes with charges $\Gamma_i = \gamma_i \times Q_i$, $i=1,2$, with $\gamma_1$ and $\gamma_2$ wrapped as in figure \ref{g123}. These can form 1/4-BPS bound states with each other. The brane configuration has the following duality invariants:
\begin{align}
 r \equiv \frac{1}{2} Q_1^2 \, , \qquad s \equiv \frac{1}{2} Q_2^2 \, , \qquad a \equiv - Q_1 \cdot Q_2 \, . 
\end{align}
The sign is chosen for consistency with conventions in later sections, but does not matter at the level of precision of this section.  
The generating function for the bound state degeneracies $\Omega(r,s,a)$ is $Z_{g=2} = 1/\Phi_{10}$ \cite{DVV}, expanded in powers of
\begin{align}
 x \equiv e^{2 \pi i \tau_{11}} \, , \qquad y \equiv e^{2 \pi i \tau_{22}} \, , \qquad u \equiv e^{2 \pi i \tau_{12}} \, ,
\end{align}
where $\tau_{ij}$ is the $g=2$ period matrix;
that is to say,
\begin{align} \label{Zg2}
 Z_{g=2}(x,y;u) = \frac{1}{\Phi_{10}(x,y;u)} = \sum_{rsa} \Omega(r,s,a) \, x^r y^s u^{\pm a} \, . 
\end{align} 
The sign ambiguity in the power of the expansion parameter $u$ depends on whether one views this as a large-$u$ or a small-$u$ expansion, and is related to ambiguities in the definition of $\Omega(r,s,a)$ due to background moduli dependence of the BPS spectrum, i.e.\ wall-crossing \cite{Sen:2007vb,Dabholkar:2007vk,Sen:2007pg,Cheng:2007ch,Banerjee:2008yu,Sen:2008ht}. There are other subtleties when the charges $Q_i$ are non-primitive \cite{Atishthree}. True to our promise, we ignore all of this here.  In the 4d $\CN=4$ low energy effective supergravity theory, these 1/4-BPS bound states are realized either as a single-center (large) black hole, or as a 2-center bound state of 1/2-BPS (small) black holes \cite{Sen:2007vb,Dabholkar:2007vk,Sen:2007pg,Cheng:2007ch,Sen:2008ht}.

We are now ready to formulate our conjecture. To this end we consider {\it three} 1/2-BPS branes with charges  $\Gamma_i = \gamma_i \times Q_i$, $i=1,2,3$, with the $\gamma_i$ wrapped as in figure \ref{g123}. The brane configuration has the following duality invariants:
\begin{align} \label{rstabcdef}
 (r,s,t) = \bigl(\tfrac{1}{2} Q_1^2,\tfrac{1}{2} Q_2^2,\tfrac{1}{2} Q_3^2 \bigr) \, , \qquad
 (a,b,c) = -\bigl( Q_1 \!\cdot\! Q_2,\,Q_2 \!\cdot \! Q_3,\,Q_3\! \cdot\! Q_1 \bigr).
\end{align}
Although for generic $Q_i$ and at generic points in the $T^2 \times K3$ moduli space, these three branes will not form BPS bound states \cite{Dabholkar:2009dq}, it is nevertheless the case that for suitable charges and on suitable subloci of the moduli space, they {\it will} form BPS bound states. We conjecture that the generating function for the bound state degeneracies $\Omega(r,s,t,a,b,c)$ is $Z_{g=3} = 1/\chi_9$, expanded in powers of
\begin{align} \label{xyzuvw}
(x,y,z) \equiv (e^{2 \pi i \tau_{11}},e^{2 \pi i \tau_{22}},e^{2 \pi i \tau_{33}}) \, , \qquad (u,v,w) \equiv (e^{2 \pi i \tau_{12}},e^{2 \pi i \tau_{23}},e^{2 \pi i \tau_{13}})
\end{align}
where $\tau_{ij}$ is the $g=3$ period matrix; that is to say,
\begin{align} \label{Zg3}
Z_{g=3}(x,y,z;u,v,w) = \frac{1}{\chi_9(x,y,z;u,v,w)} = \sum_{rstabc} \Omega(r,s,t,a,b,c) \, x^r y^s z^t u^{\pm a} v^{\pm b} w^{\pm c} \, .
\end{align}
As we will make precise in section \ref{sec:conjecture}, the sign ambiguities are again related to wall crossing ambiguities. In the 4d $\CN=4$ low energy effective supergravity theory, these bound states are realized either as a single-center  black hole, or as a 2-center bound state of 1/2-BPS and a 1/4-BPS black hole, or as a 3-center bound state of 1/2-BPS black holes. The latter includes in particular also scaling solutions. 

To state a more precise version of the conjecture and to subject it to tests, we  need a more detailed description of these black hole configurations and their wall crossing properties. We turn to this next.

\section{Black hole bound states} \label{sec:threeparticleboundstates}

In this section we study in some detail the black hole configurations corresponding to bound states of three 1/2-BPS D-branes as in fig.\ \ref{g123} on the right. The level of detail is needed because several of our tests of the conjecture use wall-crossing in an essential way, so we will need to be precise about bound state stability conditions and various signs on which these conditions depend. We build up the relevant technology in steps. We begin by making a few things in our discussion above a little bit more precise. We then discuss single-, two- and three-center black hole realizations of the bound states of interest. In the final part we review known results \cite{Denef:2007vg,Bena:2012hf,Lee:2012sc,Manschot:2013sya} about the ground state degeneracy of 3-node quiver quantum mechanics. Although these quivers do not accurately describe the bound states of interest to us, they do have similar 3-particle bound state realizations, and their degeneracies exhibit striking qualitative features suggestively similar to the degeneracies extracted from $Z_{g=3} = 1/\chi_9$.

\subsection{D-brane setup} \label{sec:Dbranesetup}

We consider again the three 1/2-BPS D-branes wrapped on cycles $\Gamma_i = \gamma_i \times Q_i$ as depicted in fig.\ \ref{g123} on the right. More precisely, denoting the horizontal and vertical 1-cycles of the $T^2$ by $A$ and $B$, oriented such that the intersection product $\langle A, B \rangle_{T^2} = +1$, we have:
\begin{align} \label{charges}
\Gamma_1 = A \times Q_1 \, , \qquad
\Gamma_2 = (B-A) \times Q_2 \, , \qquad
\Gamma_3 = - B \times Q_3 \, ,
\end{align}
where $Q_i \in H_{\rm even}(K3)$. The 1-cycles are chosen to have intersection product $\langle \gamma_i,\gamma_j\rangle_{T^2} = +1$ for all cyclically ordered pairs $(\gamma_i,\gamma_j)$. The intersection products of the $\Gamma_i$ are  
\begin{align}
a = \langle \Gamma_2,\Gamma_1 \rangle = -Q_1 \cdot Q_2 \, , \quad b =
\langle \Gamma_3,\Gamma_2 \rangle = -Q_2 \cdot Q_3 \, , \quad
c = \langle \Gamma_1,\Gamma_3 \rangle = -Q_3 \cdot Q_1  \, ,
\end{align}
as defined earlier in (\ref{rstabcdef}). We assume the charges $Q_i$ are chosen such that 
\begin{align}
a,b,c > 0 \, ,
\end{align}
and we assume there exists a locus in the $T^2 \times K3$ moduli space where the three branes are mutually supersymmetric. On general grounds \cite{Brunner:1999jq,Kachru:1999vj,Douglas:2000ah,Denef:2007vg}, supersymmetric bound states of these branes will form when moving away from this locus along suitable (though not arbitrary) directions in the moduli space. 

A simple analogous brane setup, where existence of such brane configurations is readily checked by elementary means, is given by a system of intersecting D3-branes on $T^2 \times T^2 \times T^2$ instead of $T^2 \times K3$, with $\Gamma_1 = A_1 \times A_2 \times A_3$, $\Gamma_2 = (B_1 - A_1) \times (B_2-A_2) \times (B_3-A_3)$, $\Gamma_3 = - B_1 \times B_2 \times B_3$. 
The intersection products in this case are easily computed to be $a=b=c=+1>0$. If we take the complex structure moduli of the $T^2$ factors to be $\tau_1 = \tau_2 = \tau_3 = e^{i \pi/3}$, the periods ${\cal Z}_i = \int_{\Gamma_i} \Omega$ of the holomorphic 3-form $\Omega = dz_1 \wedge dz_2 \wedge dz_3$ on $T^2 \times T^2 \times T^2$ are all equal to 1, so the branes are mutually supersymmetric at this point in the moduli space. Moving the complex structure moduli slightly away from $\tau_i = e^{i \pi/3}$ in the appropriate directions produces BPS bound states of these intersecting D-branes.

Near the locus where the three constituents are mutually supersymmetric, the low energy dynamics of the D-brane system is captured by a quiver-like $\CN=4$ supersymmetric quantum mechanics model, similar but not identical to the 3-node cyclic quiver model introduced in \cite{Denef:2007vg}. Counting BPS bound states amounts to counting the supersymmetric ground states of this system. This problem was solved for 3-node cyclic quivers with generic cubic superpotential in \cite{Denef:2007vg,Bena:2012hf,Lee:2012sc,Manschot:2013sya}. However those results are not directly applicable here. One difference is that the constituent branes necessarily have moduli in the case of interest, since there are always at least the translation and Wilson line moduli of the 1-cycles wrapped on $T^2$. This allows for more general superpotentials depending nontrivially on these moduli. For intersecting branes on $T^6$, including generalizations to more complicated wrappings with larger values of $a,b,c$, explicit expressions for the superpotential were obtained in \cite{Cremades:2003qj} in terms of theta functions. Related explicit models in IIA on $T^6$ were constructed in \cite{Chowdhury:2014yca}.

One could at this point try to deduce the appropriate microscopic description for the $T^2 \times K3$-wrapped brane systems of interest, and to identify and count the appropriate collection of BPS states directly in this description. We will not attempt this here. Instead we will consider the 4d $\CN=4$ low energy supergravity description of these bound states, and use this in combination with the interpretation of $Z_{g=1} = 1/\Delta$ and $Z_{g=2} = 1/\Phi_{10}$ as BPS counting functions to perform a number of rather nontrivial checks of our conjecture.

\def\CQ{{\cal Q}} \def\CP{{\cal P}} 

\subsection{Single-center black hole solutions}

A regular single-center BPS black hole solution with total charge $\Gamma = A \times \CQ + B \times \CP$ exists  provided \cite{Cvetic:1995bj} $\CQ^2>0$, $\CP^2>0$, $\Delta > 0$, where $\Delta$ is called the discriminant,
\begin{align}
\Delta \equiv \CQ^2 \CP^2 - (\CQ \cdot \CP)^2 \, .
\end{align}
The black hole entropy is then given by 
\begin{align}
S = \pi \sqrt{\Delta} \, . \label{SDelta}
\end{align}
When $\Delta=0$, the black hole becomes singular to leading order in the supergravity approximation, but may still have a finite size horizon when $\alpha'$-corrections are taken into account, giving rise to a small black hole. This is the case in particular for generic half-BPS charges, i.e.\ charges in the duality orbit of $\Gamma = A \times \CQ$ with $\CQ^2 \geq 0$. If $\Gamma = A \times \CQ$ with $\CQ^2=-2$, obtained for example by wrapping a D3 on a supersymmetric 2-sphere in the K3, we get an elementary BPS particle rather than a black hole. 

The total charge $\Gamma$ of the D-brane system considered above in section \ref{sec:Dbranesetup} is
\begin{align}
\Gamma = \Gamma_1 + \Gamma_2 + \Gamma_3 = A \times (Q_1-Q_2) + B \times (Q_2 - Q_3) \, ,
\end{align} 
so $\CQ = Q_1 - Q_2$ and $\CP = Q_2 - Q_3$. The corresponding duality invariants are
\begin{align}
\CQ^2 &= Q_1^2 + Q_2^2 - 2 Q_1 \cdot Q_2 = 2(r+s+a) \nonumber \\
\CP^2 &= Q_2^2 + Q_3^2 - 2 Q_2 \cdot Q_3 = 2(s+t+b) \nonumber \\
\CQ \cdot \CP &= - Q_1 \cdot Q_3 + Q_1 \cdot Q_2 + Q_2 \cdot Q_3  - Q_2^2 = c - a - b - 2 s \, . \label{CQCPCQCP}
\end{align}
Plugging in (\ref{CQCPCQCP}) and reworking things a bit, we get
\begin{align}
\Delta & = 2(ab + bc + ca) - (a^2+b^2+c^2) +4 (rs + st + tr) + 4(rb+sc+ta) \nonumber \\
&=  (\sqrt{a}+\sqrt{b}+\sqrt{c})(\sqrt{a}+\sqrt{b}-\sqrt{c})(\sqrt{a}-\sqrt{b}+\sqrt{c})(-\sqrt{a}+\sqrt{b}+\sqrt{c}) \nonumber \\
& \quad  +4 (rs + st + tr) + 4(rb+sc+ta) \, . \label{Deltaexplicit}
\end{align}
In the limit $a,b,c \gg r,s,t$, we have $\CQ^2 \gg 1$ and $\CP^2 \gg 1$, and  $\Delta>0$ provided $\sqrt{a}$, $\sqrt{b}$, $\sqrt{c}$ satisfy the triangle inequalities, i.e.\
\begin{align}
\sqrt{a}+\sqrt{b} > \sqrt{c} \, , \qquad 
\sqrt{b}+\sqrt{c} > \sqrt{a} \, , \qquad
\sqrt{c}+\sqrt{a} > \sqrt{b} \, . \label{squareroottriangineq}
\end{align}
It can be checked that this remains a sufficient condition for existence of the BPS black hole even when the $a,b,c$ are not parametrically larger than $r,s,t$.

\subsection{Two-center solutions} \label{sec:twocentersol}

Under suitable conditions, 2-center black hole bound states of the same total charge $\Gamma$ exist. In general there may be a huge number of different ways of splitting up $\Gamma = \Gamma_A + \Gamma_B$ to form a 2-center bound state with charges $\Gamma_A,\Gamma_B$, but here we will only consider charges obtained by merging two out of the three constituent charges $\Gamma_1,\Gamma_2,\Gamma_3$ into a black hole, and binding this to the third one as a 2-center bound state. Consider first the split 
\begin{align}
\Gamma_A=\Gamma_1+\Gamma_2 = A \times (Q_1-Q_2) + B \times Q_2,  \qquad 
\Gamma_B=\Gamma_3 = - B \times Q_3 \, . 
\end{align}
The quadratic invariants for $\Gamma_A$ are
\begin{align}
\CQ_A^2 = (Q_1-Q_2)^2 = 2(r+s+a) \, , \qquad \CP_A^2 = 2s \, , \qquad \CQ_A \cdot \CP_A = -(a+2s) \, ,
\end{align}
so its discriminant is
\begin{align}
\Delta_A = \CQ_A^2 \CP_A^2 - (\CQ_A \cdot \CQ_B)^2 = 4 r s - a^2 \, .
\end{align}
Thus a regular single-center black hole of charge $\Gamma_A$ exists provided 
\begin{align}
r,s,t > 0 \, , \qquad 4 r s > a^2 \, .
\end{align} 
Its entropy is $S_A = \pi \sqrt{4rs-a^2}$.

These two centers may form a bound state with equilibrium separation given by the BPS constraint \cite{Denef:2000nb,Bates:2003vx}
\begin{align} \label{twocenterposition}
\frac{\langle \Gamma_B,\Gamma_A \rangle}{|x_B-x_A|} = \theta_B = -\theta_A \, .
\end{align}
The constants $\theta_B$, $\theta_A$ are determined by the charges and background moduli. In ${\cal N}=2$ language, they are more specifically expressed in terms of the central charges $Z_A$ and $Z_B$, as $\theta_B = 2 \, {\rm Im}\bigl( e^{-i\alpha} Z_B \bigr)$, $\theta_A = 2 \, {\rm Im}\bigl( e^{-i\alpha} Z_A \bigr) = - \theta_B$, where $e^{i\alpha} \equiv \frac{Z}{|Z|}$, $Z \equiv Z_A + Z_B$. Since distances are positive, the bound state can only exist if $\langle \Gamma_B,\Gamma_A \rangle \theta_B > 0$. In the case at hand, $\theta_B=\theta_3 = {\rm Im}\bigl( e^{-i\alpha} Z_3 \bigr)$ and $\langle \Gamma_B,\Gamma_A \rangle = Q_3 \cdot (Q_1-Q_2) = b-c$, so the existence condition becomes $(b-c) \theta_3 > 0$. 
The BPS degeneracy associated with this configuration is $\Omega = |\langle \Gamma_B,\Gamma_A \rangle| \, \Omega_A \, \Omega_B$, 
where $\Omega_A$, $\Omega_B$ are the degeneracies of the two centers and $|\langle \Gamma_B,\Gamma_A \rangle| = |b-c|$ is an electromagnetic intrinsic angular momentum degeneracy \cite{Denef:2002ru}. 

The other charge splittings, namely $\Gamma_A = \Gamma_2 + \Gamma_3$, $\Gamma_B=\Gamma_1$ and $\Gamma_A=\Gamma_3+\Gamma_1$, $\Gamma_B=\Gamma_2$ can be treated analogously. To summarize, we get the following existence conditions and degeneracies for the three possible two-center bound states under consideration:
\begin{align}
(\Gamma_1+\Gamma_2,\Gamma_3): && r,s,t > 0 \, , \,\, 4rs > a^2 \, , \,\, (b-c) \theta_3 > 0  &\quad \leadsto& \Omega = |b-c| \, \Omega_{1+2} \, \Omega_3  \nonumber \\
(\Gamma_2+\Gamma_3,\Gamma_1): && r,s,t > 0 \, , \,\, 4st > b^2 \, , \,\, (c-a) \theta_1 > 0  &\quad \leadsto& \Omega = |c-a| \, \Omega_{2+3} \, \Omega_1 \nonumber \\
(\Gamma_1+\Gamma_2,\Gamma_3): && r,s,t > 0 \, , \,\, 4rs > c^2 \, , \,\, (a-b) \theta_2 > 0  &\quad \leadsto& \Omega = |a-b| \, \Omega_{3+1} \, \Omega_2  \label{Omfacttwo}
\end{align} 
Marginal cases in which some of these inequalities are relaxed to equalities may exist as well, as discussed under (\ref{SDelta}).  Note that in the limit $a,b,c \gg r,s,t$, none of these 2-center solutions exist at all. 
The exact single-center degeneracies $\Omega_i$ and $\Omega_{j+k}$ can be obtained from the generating functions $Z_{g=1}=1/\Delta$ and $Z_{g=2}=1/\Phi_{10}$, as in 
(\ref{Zg1}) and (\ref{Zg2}). 

\subsection{Three-center solutions} \label{sec:threecentersol}

Likewise, three-center bound states with center charges $\Gamma_1,\Gamma_2,\Gamma_3$ may exist. The BPS position constraints generalizing (\ref{twocenterposition}) to three centers are
\begin{align}
 \frac{\langle \Gamma_1,\Gamma_3 \rangle}{|x_1-x_3|} 
 + \frac{\langle \Gamma_1,\Gamma_2 \rangle}{|x_1-x_2|}  = \theta_1 \, , \quad
\frac{\langle \Gamma_2,\Gamma_1 \rangle}{|x_2-x_1|}
+ \frac{\langle \Gamma_2,\Gamma_3 \rangle}{|x_2-x_3|} = \theta_2 \, , \quad
\frac{\langle \Gamma_3,\Gamma_2 \rangle}{|x_3-x_2|}
+ \frac{\langle \Gamma_3,\Gamma_1 \rangle}{|x_3-x_1|}  = \theta_3 \, , \nonumber
\end{align} 
that is
\begin{align}
\frac{c}{|x_1-x_3|} - \frac{a}{|x_1-x_2|} = \theta_1 \, , \quad
\frac{ a }{|x_2-x_1|} -\frac{b}{|x_2-x_3|} 
= \theta_2 \, , \quad
\frac{b}{|x_3-x_2|}
-\frac{c}{|x_3-x_1|} 
= \theta_3 \, \nonumber \, .
\end{align} 
Here $\theta_i = 2 \, {\rm Im}(e^{-i \alpha} Z_i)$, $e^{i \alpha} = \frac{Z}{|Z|}$, $Z=Z_1+Z_2+Z_3$. Note that this implies $\theta_1+\theta_2+\theta_3 = 0$, so summing up the three equations above just gives $0=0$.  

These equations do not always have solutions. Assume for example $\theta_3>0$, $\theta_1 < 0$. Then the first equation implies $\frac{a}{|x_1-x_2|} > \frac{c}{|x_1-x_3|}$ and the second equation implies $\frac{b}{|x_3-x_2|} > \frac{c}{|x_1-x_3|}$. Combining these two inequalities implies $a+b > \frac{|x_1-x_2|+|x_3-x_2|}{|x_1-x_3|} \, c \geq c$, where  the last inequality follows from the the triangle inequalities for the $(x_1,x_2,x_3)$-triangle. Therefore if $a+b \leq c$ and $\theta_3 > 0$, $\theta_1 < 0$, the 3-center bound state does not exist. 

On the other hand, if $a+b > c$, $b+c >  a$, $c+a >  b$, that is to say, if the intersection products $(a,b,c)$ satisfy the triangle inequalities, a branch of 3-center solutions always exists. This branch is connected to so-called scaling solutions \cite{Denef:2007vg,Denef:2002ru,Bena:2006kb}, consisting of configurations for which the centers approach each other arbitrarily closely in coordinate space, $|x_i-x_j| \to 0$. In this limit, the position constraints reduce to the scale-invariant equations
\begin{align}
\frac{a}{|x_1-x_2|} = \frac{b}{|x_2-x_3|} = \frac{c}{|x_3-x_1|} \, .
\end{align} 
Consistency with the triangle inequalities for the $(x_1,x_2,x_3)$-triangle then requires that $a,b,c$ themselves satisfy the triangle inequalities, i.e.\
\begin{align}
a+b > c, \qquad b+c > a, \qquad c+a > b \, . \label{triangleineq}
\end{align}
(Marginal cases in which an inequality becomes equality require a more careful discussion, as the earlier analysis of the $\theta_3>0$, $\theta_1<0$ case illustrates, but we will skip this.)

Although the coordinate size of these configurations goes to zero in the limit, their physical size remains finite in the full supergravity solution \cite{Bena:2006kb}. In fact, in the scaling limit, the solution becomes indistinguishable to a distant observer from a single-center BPS black hole solution of total charge $\Gamma=\Gamma_1+\Gamma_3+\Gamma_3$. Consistency thus requires that the single-center discriminant $\Delta$ as given in (\ref{Deltaexplicit}) is positive whenever the triangle inequalities (\ref{triangleineq}) are satisfied. Happily, this is the case, as in general (\ref{triangleineq}) implies (\ref{squareroottriangineq}). The implication does not run the other way, so the existence of scaling solutions implies the existence of single-center solutions, but the converse is not true in general. 

If the triangle inequalities are {\it not} satisfied, scaling solutions do not exist, and by tuning the background moduli close to values where the three central charge phases line up (so $\theta_i \to 0$), the centers can be taken to be arbitrarily well-separated. In this case the BPS degeneracy of the configuration can be determined from wall crossing arguments \cite{Denef:2007vg} or by direct quantization of the system \cite{deBoer:2008zn,Manschot:2010qz,Manschot:2013sya,Manschot:2014fua}. Explicitly, 
\begin{align} \label{Omfactthree}
\Omega = \Omega_c \, \Omega_1 \, \Omega_2 \, \Omega_3 \, , 
\end{align}
where $\Omega_c$ is the ``configurational'' degeneracy (discussed below) and the $\Omega_i$ are the BPS degeneracies of the 1/2-BPS centers $\Gamma_i$. The exact single-center degeneracies $\Omega_i$  can be obtained from the generating functions $Z_{g=1}=1/\Delta$ as in 
(\ref{Zg1}).   

The configurational factor $\Omega_c$ is most easily obtained from wall crossing. Assuming $a+b \leq c$, we know from our earlier discussion above that $\Omega_c=0$ in the moduli space chamber $\theta_1>0$, $\theta_3<0$. When passing to other chambers, this may jump to nonzero values. However such jumps can only happen at walls where the size of a bound state diverges, i.e.\ when one of the centers is pushed out to or pulled in from infinity. If the first center goes to infinity, then $R \equiv |x_1-x_2| \approx |x_1-x_3|  \to \infty$ while $r \equiv |x_2-x_3|$ remains finite. From the position constraint equations it then follows that $\frac{c-a}{R} \approx \theta_1$, $\frac{b}{r} \approx -\theta_2 \approx \theta_3$. Hence jumps of this kind occur when $\theta_3>0$, $\theta_2<0$ while $\theta_1$ passes through zero, with the bound state existing on the side with $(c-a) \theta_1 > 0$. Since we assumed $a+b \leq c$, we have in particular $c-a>0$ so the bound state exists when $\theta_1>0$. Near the transition, $\Gamma_1$ is loosely bound to a tighter bound state of $\Gamma_2$ and $\Gamma_3$. The latter has degeneracy $|\langle \Gamma_3,\Gamma_2\rangle|=b$, and binding this $(2+3)$-atom to the first center multiplies this degeneracy by a factor $|\langle \Gamma_1,\Gamma_2+\Gamma_3\rangle| = c-a$, resulting in a total configurational degeneracy $\Omega_c = b(a-c)$. Thus we conclude that in the region $\theta_1 > 0$, $\theta_2 < 0$, $\theta_3 > 0$, we have $\Omega_c = b(a-c)$. Similar arguments can be used to determine $\Omega_c$ in all chambers, as well as for the other possible violations of the triangle inequalities. We summarize the results for $\Omega_c$ in all of these cases in the following table:
\begin{center} 
\begin{tabular}{|ccc|ccc|} 
\hline 
$\theta_1$&$\theta_2$&$\theta_3$&$a+b \leq c$&$b+c \leq a $& $c+a \leq b$ \\
\hline
$-$ & $\pm$ & $+$ & $\Omega_c = 0$ & $b(a-c)$ & $a(b-c)$ \\ 
$+$ & $-$ & $\pm$ & $b(c-a)$ & $0$ & $c(b-a)$ \\ 
$\pm$ & $+$ & $-$ & $a(c-b)$ & $c(a-b)$ & $0$ \\ 
\hline
\end{tabular} 
\end{center}
Notice that the {\it difference} $\Delta \Omega_c$ between these rows is always the same, independent of which triangle inequality is violated, as dictated by the wall crossing formula. For example the difference between rows 2 and 1 is 
\begin{align} \label{Omegacjump}
\Omega_{c,2} - \Omega_{c,1} = b(c-a)
\end{align}
for all three cases. Moreover, since the jumps are entirely determined by bound states of divergent size, as opposed to scaling solutions, the same $\Delta \Omega$ can be expected to apply even when the triangle inequalities {\it are} satisfied. This is borne out by explicit computations in microscopic quiver models \cite{Denef:2007vg,Bena:2012hf,Lee:2012sc,Manschot:2013sya}. 

To get the total degeneracy when scaling solutions {\it do} exist, i.e.\ when $(a,b,c)$ satisfy the triangle inequalities, one has to take into account condensation of light open strings stretched between the constituent branes in the scaling regime, also known as the ``Higgs branch'' of the system. A simple 3-node cyclic quiver model sharing this feature was considered in \cite{Denef:2007vg} and further analyzed in \cite{Bena:2012hf,Lee:2012sc,Manschot:2013sya}. As mentioned before at the end of section \ref{sec:Dbranesetup}, there is no reason to expect this to be a quantitatively accurate model for the bound states of interest to us. However it should nevertheless give a reasonable model for at least some qualitative features of the degeneracies. A generating function counting BPS states of this model was found in \cite{Bena:2012hf}. In the chamber chamber $\theta_1<0$, $\theta_2<0$, $\theta_3>0$, it is given by
\begin{align} 
\sum_{a,b,c} \Omega(a,b,c) \, u^a v^b w^c  =  \frac{uv(1-uv)}{(1-u)^2(1-v)^2(1-uv-vw-wu + 2 \, uvw)} \, .  
\end{align} 
For example for $(a,b)=(5,9)$, the degeneracies are given by
\begin{center} \footnotesize
\begin{tabular}{|r|rrrrrrrrrrrrrrrr|} 
\hline 
c&0&1&2&3&4&5&6&7&8&9&10&11&12&13&14&15 \\
\hline
$\Omega(5,9,c)$&45& 40& 35& 30& 25& 20& 141& --578& 1583& --2556& 2685& --1650& 495& 0& \
0& 0 \\
 \hline
\end{tabular}
\end{center}
This agrees with the previous table for the range in which the triangle inequalities are {\it not} satisfied, $c + 5 \leq 9$ and $5 + 9 \leq c$. When they {\it are} satisfied,  the degeneracies become exponentially large. Asymptotically for large $a$, $b$, $c$ \cite{Bena:2012hf}, 
\begin{align} \label{threenodequiverasymptotics}
\Omega(\alpha N,\beta N,\gamma N) \sim \left(\frac{\alpha^\alpha \beta^\beta \gamma^\gamma \, 2^{\alpha+\beta+\gamma}}{(\alpha+\beta-\gamma)^{\alpha+\beta-\gamma} (\alpha+\gamma-\beta)^{\alpha+\gamma-\beta} (\beta+\gamma-\alpha)^{\beta+\gamma-\alpha} }\right)^N \, .
\end{align}
In particular when $a=b=c=N$, this becomes $\Omega(N,N,N) \sim 2^{3N}$. 

\section{A more precise conjecture} \label{sec:conjecture}

From the discussion above it is clear that the 3-constituent bound states degeneracies conjecturally counted by $1/\chi_9$ depend on the background moduli, and considerably more so than the 2-constituent bound states counted by $1/\Phi_{10}$. More specifically, the spectrum depends on the signs of the parameters $\theta_1,\theta_2,\theta_3$ introduced above. In the following section we will provide evidence that these wall crossing ambiguities are related to the sign ambiguities in our conjecture as formulated in (\ref{Zg3}). Based on this evidence, a more precise version of the conjecture appears to be
\begin{align} \label{Zg3b}
 Z_{g=3}(x,y,z;u,v,w) = \frac{1}{\chi_9(x,y,z;u,v,w)} = \sum_{rstabc} \Omega(r,s,t,a,b,c) \, x^r y^s z^t u^{\rho a} v^{\sigma b} w^{\tau c} \, ,
\end{align}
where $(\rho,\sigma,\tau)$ are the following $\theta$-dependent signs:
\begin{center} 
\begin{tabular}{|ccc|ccc|} 
\hline 
$\theta_1$&$\theta_2$&$\theta_3$&$\rho$&$\sigma$& $\tau$ \\
\hline
$-$ & $\pm$ & $+$ & $+$ & $+$ & $-$ \\ 
$+$ & $-$ & $\pm$ & $-$ & $+$ & $+$ \\ 
$\pm$ & $+$ & $-$ & $+$ & $-$ & $+$ \\ 
\hline
\end{tabular}
\end{center}
An ambiguity still implicit in (\ref{Zg3b}) is that the Taylor expansion of $Z_{g=3}$ depends on the order in which $u$, $v$ and $w$ are expanded. Equivalently, if the coefficients are extracted by contour integration, the result depends on the contour, and more specifically on the relative sizes of $|u|$, $|v|$ and $|w|$. In the above it is understood that the variable with the negative sign is expanded last. So for example if $\theta_1<0$, $\theta_3>0$ (row 1), we first expand $u$, $v$ around zero (the order does not matter in this case), and next $w$ around zero. Alternatively this corresponds to picking a contour with $|u|,|v| \ll |w| < 1$. 

Another way of phrasing the above table is that in each chamber we have a particular ordering of $\{\Gamma_1,\Gamma_2,\Gamma_3 \}$, to wit
\begin{center} 
\begin{tabular}{|ccc|c|} 
\hline 
$\theta_1$&$\theta_2$&$\theta_3$& ordering \\
\hline
$-$ & $\pm$ & $+$ & $\Gamma_1 < \Gamma_2 < \Gamma_3$  \\ 
$+$ & $-$ & $\pm$ & $\Gamma_2 < \Gamma_3 < \Gamma_1$ \\ 
$\pm$ & $+$ & $-$ & $\Gamma_3 < \Gamma_1 < \Gamma_2$  \\ 
\hline
\end{tabular}
\end{center}
Then the power $m_{ij}$ of the off-diagonal $e^{2 \pi \tau_{ij}}$ in the expansion is fixed by
\begin{align}
m_{ij} = \langle \Gamma_j,\Gamma_i \rangle \quad \mbox{ if } \quad \Gamma_j > \Gamma_i .
\end{align}
The Taylor expansion order (or integration contour) is likewise specified by this ordering, with the expansion order determined by the induced pair ordering. For example if $\Gamma_2 < \Gamma_3 < \Gamma_1$, we first expand in $v=e^{2 \pi i \tau_{23}}$, then in $w=e^{2 \pi i \tau_{31}}$, and finally in $u=e^{2 \pi i \tau_{12}}$.

\section{Tests at leading order in the $q$-expansion of $1/\chi_9$} \label{sec:leading-q-examples}

Recalling the definition (\ref{xyzuvw}) of the expansion parameters, let us explore the conjecture for the leading term in the $q$-expansion, i.e.\ the small $(x,y,z)$-expansion of $1/\chi_9$, that is
\begin{align}
\frac{1}{\chi_9(x,y,z;u,v,w)} = x^{-1} y^{-1} z^{-1} Z(u,v,w) + \cdots \, , 
\end{align}
where
\begin{align}
Z(u,v,w) &= \frac{u v w}{(1-u)(1-v)(1-w)\sqrt{P(u,v,w)}}  \nonumber \\
P(u,v,w) &= u^2+v^2+w^2-2(uv+vw+wu)-2\,uvw(u+v+w-4)+u^2v^2w^2 \nonumber \\
&=(\sqrt{u}+\sqrt{v}+\sqrt{w}+\sqrt{u v w})(\sqrt{u}-\sqrt{v}-\sqrt{w}+\sqrt{u v w}) \nonumber \\
& \quad \times (-\sqrt{u}+\sqrt{v}-\sqrt{w}+\sqrt{u v w}) (-\sqrt{u}-\sqrt{v}+\sqrt{w}+\sqrt{u v w}) \, .
\end{align}
Note that $Z(u,v^{-1},w^{-1}) = Z(u,v,w)$ and cyclic permutations thereof, but $Z(u,v,w^{-1}) \neq Z(u,v,w)$. According to the conjecture, the expansion of $Z$ in powers of $u,v,w$ should count bound states in the $(r,s,t) = (-1,-1,-1)$ sector of our D-brane setup. As recalled under (\ref{Zg1}), this means that each of the constituent 1/2-BPS D-branes wraps a rigid cycle $Q_i$. According to (\ref{Omfacttwo}), there are no 2-center bound states to consider in this sector. Therefore the degeneracies predicted by $1/\chi_9$ should count the three-particle bound states discussed in \ref{sec:threecentersol}. 

We asserted in section \ref{sec:conjecture} that Taylor expanding $Z$ about $(u,v,w)=(0,0,0)$ depends on the order in which we are expanding, or equivalently on the contour we use to extract the coefficients.  For example, if we first expand in $u$, then in $v$, and finally in $w$, we get, up to cubic order in $u$, $v$ and up to zeroth\footnote{All terms of positive order in $w$ have the same coefficient as the corresponding terms of zeroth order in $w$. For example the terms of order $O(u^2 v^3 w^n)$, $n \geq 0$, are $u^2 v^3(6+6 w + 6 w^2 + 6 w^3 + 6 w^4 + \cdots)$.} order in $w$:
\begin{align}
Z(u,v,w) &= u \bigl(v + v^2 (w^{-1}+ 2) + v^3 (w^{-2} + 2 \, w^{-1} + 3) \bigr) \nonumber \\ 
&\quad + u^2 \bigl(v (w^{-1} + 2) + v^2 (4 \, w^{-2} + 2 \, w^{-1} + 4) + 
   v^3 (9 \, w^{-3} + 2 \, w^{-2} + 4 \, w^{-1} + 6) \bigr) \nonumber \\
&\quad + 
u^3 \bigl(v (w^{-2} + 2\, w^{-1} + 3) + 
   v^2 (9 \,w^{-3} + 2 \,w^{-2} + 4 \,w^{-1} + 6) \nonumber \\
&\quad \quad \quad \,\, + 
   v^3 (36 \,w^{-4} - 18 \,w^{-3} + 12 \,w^{-2} + 6\, w^{-1} + 9) \bigr) + \cdots \, .
\end{align}
While this is symmetric under exchange $u \leftrightarrow v$, it is evidently not symmetric under exchange of $u \leftrightarrow w$, nor under exchange of $u \leftrightarrow w^{-1}$. 

Denote the coefficient of $u^a v^b w^{-c}$ in this $|u|,|v| \ll |w| \ll 1$ expansion by $\Omega(a,b,c)$. An alternative expansion is obtained by first expanding in $u$ and $w$ and then in $v$, or equivalently $|u|, |w| \ll |v| \ll 1$. The analog of $\Omega(a,b,c)$ is now the coefficient of $u^a w^c v^{-b}$. Denote this coefficient by $\Omega'(a,b,c)$. Finally we can consider $|v|,|w| \ll u \ll 1$, with coefficients $\Omega''(a,b,c)$. The coefficients $\Omega(a,b,c)$, $\Omega'(a,b,c)$ and $\Omega''(a,b,c)$ are almost the same, but not quite. In the table below the coefficients are listed for $(a,b)=(5,9)$:

\begin{center} \tiny
\begin{tabular}{|r|rrrrrrrrrrrrrrrrr|} 
\hline 
c&0&1&2&3&4&5&6&7&8&9&10&11&12&13&14&15&16 \\
\hline
$\Omega(5,9,c)$&45& 40& 35& 30& 25& 20& 15891& --144638& 569633& --1210896& 1451475& \
--925650& 245025& 0& 0& 0& 0 \\
$\Omega'(5,9,c)$&0& 0& 0& 0& 0& 0& 15876& --144648& 569628& --1210896& 1451480& --925640& 
245040& 20& 25& 30& 35 \\
$\Omega''(5,9,c)$& 0& 4& 8& 12& 16& 20& 15900& --144620& 569660& --1210860& 1451520& --925596& 245088& 72& 81& 90& 99 \\
$\Omega'-\Omega$& --45& --36& --27& --18& --9& 0& 9& 18& 27& 36& 45& 54& 63& 72& 81& 90& 99 \\
$\Omega''-\Omega$&--45& --40& --35& --30& --25& --20& --15& --10& --5& 0& 5& 10& 15& 20& 25& 30&35 \\
\hline
\end{tabular}
\end{center}
This exhibits a few notable features. First, the coefficients become exponentially large in some range: roughly when $(a,b,c)$ satisfy the triangle inequalities. Second, although in the exponential regime, $\Omega$, $\Omega'$, $\Omega''$ are practically the same, there is a persistent difference throughout. In fact the difference always equals 
\begin{align}
\Omega'(a,b,c) - \Omega(a,b,c) = a(c-b) \, , \qquad
\Omega''(a,b,c) - \Omega(a,b,c) = b(c-a) \, .
\end{align}
Third, $a+b \leq c$ implies $\Omega(a,b,c)=0$ and $c+a \leq b$ implies $\Omega''(a,b,c)=0$. Similarly, though not visible in this example, $b+c \leq a$ implies $\Omega'(a,b,c)=0$. We see that this exactly matches the non-triangle degeneracies  and wall crossing relations discussed in our analysis of 3-center configurations, if and only if we make the chamber/expansion identifications proposed in section \ref{sec:conjecture}. 

Here, we have just discussed some simple examples.
More general proofs and generalizations of our checks to higher order terms in the $q$-expansion can be carried out, and will appear in \cite{Zimo}. 
One can also obtain the large $a$, $b$, $c$ asymptotics of $\Omega(a,b,c)$, similar to (\ref{threenodequiverasymptotics}), which
we reproduce here (see \cite{Zimo} for further details):
\begin{equation} \label{asymptoticformula}
\Omega(\alpha N,\beta N,\gamma N)
\sim \left(\frac{(\alpha+\beta+\gamma)^{\alpha+\beta+\gamma}}{(\alpha+\beta-\gamma)^{\alpha+\beta-\gamma}(\alpha+\gamma-\beta)^{\alpha+\gamma-\beta}(\beta+\gamma-\alpha)^{\beta+\gamma-\alpha}}\right)^N.
\end{equation}
This asymptotic formula is valid provided $\alpha$, $\beta$ and $\gamma$ satisfy the triangle inequalities $\alpha+\beta > \gamma, \beta + \gamma > \alpha, \gamma+ \alpha> \beta$.
The degeneracy is exponentially large if these inequalities are satisfied and $N\gg 1$. Physical consistency requires that the degeneracy can only become exponentially large if scaling solutions exist. According to (\ref{triangleineq}), scaling solutions exist provided $(a,b,c)=N(\alpha,\beta,\gamma)$ satisfy the linear triangle inequalities, in striking agreement with what we find here from the behavior of the coefficients of $Z$. 

The expression (\ref{asymptoticformula}) is reminiscent of (\ref{threenodequiverasymptotics}), but is nevertheless different. In particular when $a=b=c=N$, the above becomes $\Omega(N,N,N) \sim 3^{3N}$, in contrast to $\Omega(N,N,N) \sim 2^{3N}$ one obtains from (\ref{threenodequiverasymptotics}). Still, the similarity is rather suggestive. Presumably the difference is due to differences in the microscopic model describing the bound states, in particular the superpotential. It would be very interesting to show this explicitly.

\begin{figure}[h]
\begin{center}
 \includegraphics[width=0.4\textwidth]{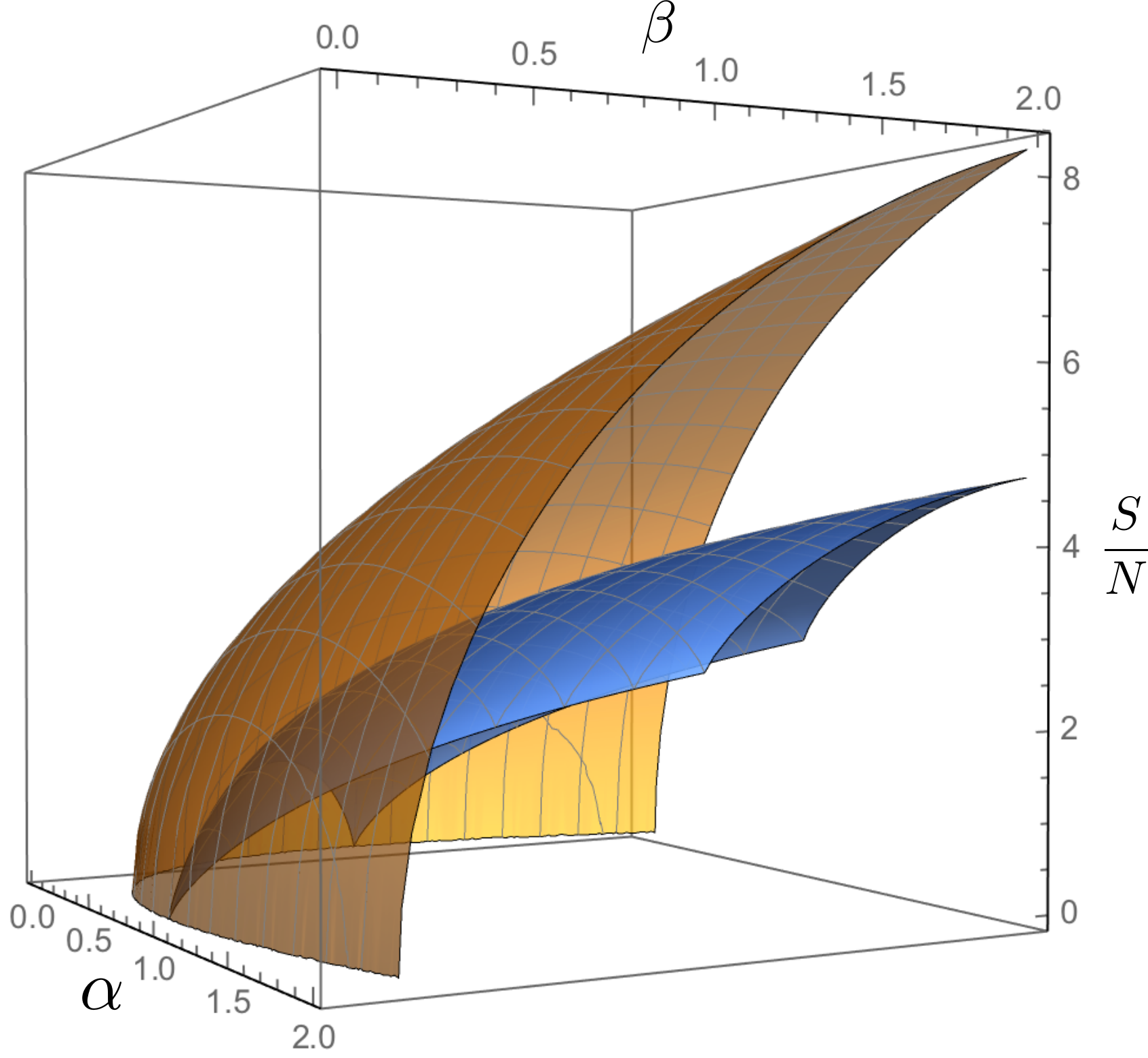}
\end{center}
\caption{Single center black hole entropy $S(\alpha N,\beta N,N)$ (orange) is always larger than 3-center entropy $\log \Omega(\alpha N,\beta N,N)$ (blue).}
\label{entropy-comparison}
\end{figure}

Either way, according to our conjecture, the asymptotic formula (\ref{asymptoticformula}) should give the degeneracy of the Higgs branch of 3-particle scaling solutions in our setup, analogous to the quiver model asymptotics (\ref{threenodequiverasymptotics}). This degeneracy should correspond to a fraction of the total degeneracy of charge $\Gamma$. It can only be a fraction because there are many other ways of splitting up $\Gamma$, and the total degeneracy should sum over all of those. Hence the predicted $\log \Omega(a,b,c)$ should be bounded above by the single-center horizon area entropy. Another argument is the holographic principle: since scaling solutions can be squeezed into a region with surface area equal to the single-center black hole horizon area $A$, their entropy must be bounded above by $A/4G = S_{\rm BH}$:
\begin{align}
\log \Omega(a,b,c) \leq S_{BH} = \pi \sqrt{\Delta(a,b,c)} \, ,
\end{align} 
with $\Delta(a,b,c)$ given by (\ref{Deltaexplicit}) with $r=s=t=-1$ and $a,b,c \gg 1$ satisfying the triangle inequalities. For example for $a=b=c=N$, this translates to the requirement $3 \log 3 N \leq \pi \sqrt{3} N$, which is indeed satisfied as $3 \log 3 \approx 3.3$ and $\pi \sqrt{3} \approx 5.4$. Happily, the inequality persists for all values $(a,b,c)$, as can be seen from fig.\ \ref{entropy-comparison} combined with some simple considerations of asymptotics. 

\section{Higher order test and the appearance of $1/\Phi_{10}$} \label{sec:higher-order}

In section \ref{sec:leading-q-examples} we gave examples illustrating that expansion ambiguities reproduce precisely the expected wall crossing formulae for 3-center configurations in the $(r,s,t)=(-1,-1,-1)$ sector. However this sector is quite insensitive to the detailed geometry of $K3 \times T^2$, because the constituents in this case are rigid and do not probe the internal space geometry. In particular, the same wall crossing formulae would be obtained for the simple 3-node quiver model with cubic superpotential (which however does not reproduce the correct degeneracies in the scaling regime). So this test, although it passes a number of nontrivial self-consistency checks, is still fairly weak in terms of singling out specifically $K3 \times T^2$ as the relevant compactification manifold. 

To unambiguously see the fingerprints of $K3 \times T^2$ in wall crossing formulae, we need to consider larger values of $(r,s,t)$ as well as values of $(a,b,c)$ such that {\it two}-center black hole bound states may form and play a role in wall crossing, not just three-center ones. Indeed for larger values of $(r,s,t)$, the individual $\Omega_i$ appearing in 3-center degeneracy formulae such as (\ref{Omfactthree}) are counted by the 1/2-BPS partition function $1/\Delta$ for $K3 \times T^2$ compactifications, and the $\Omega_{i+j}$ appearing in 2-center degeneracy formulae such as (\ref{Omfacttwo}) are counted by the 1/4-BPS partition function $1/\Phi_{10}$. Although $1/\Delta$ appears in many contexts in string theory, $1/\Phi_{10}$ is pretty much a smoking gun for $K3 \times T^2$ specifically. 

Recall that
\begin{align} \label{oneoverDelta}
\frac{1}{\Delta(x)} =  \frac{1}{x} +24 +324 \, x+3200 \, x^2
+ 25650 \, x^3 
+ \cdots
\end{align}
and
{\begin{align} 
\frac{1}{\Phi_{10}(y,z;v)} &=\frac{1}{y z}(v+2 \, v^2+\cdots)+\left(\frac{1}{y}+\frac{1}{z}\right)(2 + 24 \, v + 48 \, v^2+\cdots)\nonumber\\
&+\left(\frac{z}{y}+\frac{y}{z}\right)\left( \frac{3}{v} +48+ 327 \, v + 648\,  v^2+\cdots\right)+\left(\frac{24}{v} + 600\,  v + 1152 \, v^2+\cdots\right)\nonumber\\
&+(y+z)\left( \frac{48}{v^2} + \frac{600}{v} -648 + 8376\,  v + 15600\,  v^2+\cdots\right)\nonumber\\
&+y z\left( \frac{327}{v^3} - \frac{648}{v^2} + \frac{25353}{v} -50064+ 130329 \, v + 209304\,  v^2+\cdots\right)+\cdots
\end{align}}

Below we bring focus to a number of concrete examples evident in these expansions which strikingly confirm the $K3 \times T^2$ expectations, including both the appearance of $1/\Delta$ and $1/\Phi_{10}$: 
\begin{center} \tiny
\captionof{table}{Degeneracies in the $(-1,-1,0)$ sector}
\begin{tabular}{|r|rrrrrrrrrrrrr|} 
\hline 
c&1&2&3&4&5&6&7&8&9&10&11&12&13 \\
\hline
$\Omega(5,7,c)$&720& 600& --1870& 74760& --631504& 2698992& -6550056& 9308496& --7467552& 2968704& -387000\ & 0& 0 \\
$\Omega'(5,7,c)$& 0&0&--2350&74400&--631744& 2698872& -6550056&9308616&--7467312&2969064&--386520& 600& 720 \\
$\Omega''(5,7,c)$&48&96&--2206&74592&--631504&2699160&--6549720&9309000&--7466880
&2969544& --385992&1176&1344 \\
$\Omega'-\Omega$& --720& --600& --480& --360& --240& --120& 0& 120& 240& 360 &480& 600& 720  \\
$\Omega''-\Omega$&--672& --504& --336& --168& 0& 168& 336& 504& 672& 840& 1008& 1176& 1344 \\
\hline
\end{tabular}
\end{center}
So in this sector, the relations of these degeneracies are
\begin{equation}
\Omega'(a,b,c)-\Omega(a,b,c)=24 \times a(c-b) \qquad \Omega''(a,b,c)-\Omega(a,b,c)=24 \times b(c-a) 
\end{equation}
\begin{center} \tiny
\captionof{table}{Degeneracies in the $(-1,-1,1)$ sector}
\begin{tabular}{|r|rrrrrrrrrr|} 
\hline 
c&2&3&4&5&6&7&8&9&10&11 \\
\hline
$\Omega(4,6,c)$&5552&--20106&399488& --1967682& 4729920& --5762306& 3336160& --675690& 37824& 0 \\
$\Omega'(4,6,c)$& 368&--23994&396896&--1968978&4729920&--5761010& 3338752&--671802&43008&6480 \\
$\Omega''(4,6,c)$&1664&---22050&399488&--1965738&4733808&--5756474&3343936&--665970
&49488& 13608 \\
$\Omega'-\Omega$& --5184& --3888& --2592& --1296& 0& 1296& 2592& 3888& 5184& 6480\\
$\Omega''-\Omega$&--3888& --1944& 0& 1944& 3888& 5832& 7776& 9720& 11664& 13608 \\
\hline
\end{tabular}
\end{center}
In this sector, the relations of these degeneracies are
\begin{equation}
\Omega'(a,b,c)-\Omega(a,b,c)=324 \times a(c-b) \qquad \Omega''(a,b,c)-\Omega(a,b,c)=324 \times b(c-a) 
\end{equation}
\begin{center} \tiny
\captionof{table}{Degeneracies in the $(-1,0,1)$ sector}
\begin{tabular}{|r|rrrrrrrr|} 
\hline 
c&3&4&5&6&7&8&9&10\\
\hline
$\Omega(3,5,c)$&--407472&4448168&--13869776& 19957200&--12621568& 3200704& --200250&  0 \\
$\Omega'(3,5,c)$& --454128&4424840&--13869776&19980528&--12574912&3270688& --106938&116640 \\
$\Omega''(3,5,c)$&--407472&4487048&--13792016&20073840&--12466048&3395104&33030
&272160 \\
$\Omega'-\Omega$& --46656& --23328&0&23328& 46656& 69984& 93312& 116640\\
$\Omega''-\Omega$&0&38880& 77760& 116640& 155520& 194400& 233280& 272160 \\
\hline
\end{tabular}
\end{center}
In this sector, the relations of these degeneracies are
\begin{equation}
\Omega'(a,b,c)-\Omega(a,b,c)=24\times 324 \times a(c-b) \qquad \Omega''(a,b,c)-\Omega(a,b,c)=24\times 324 \times b(c-a) 
\end{equation}

Note in the second and third tables, we did not include some small values of $c$, because wall crossing at these charges is captured by $\Phi^{-1}_{10}\times\Delta^{-1}$ instead of $\Delta^{-1}\times\Delta^{-1}\times\Delta^{-1}$: in this  regime, the bound state decay products include a 1/4-BPS black hole. Here are some numerical examples.
Consider $(a,b,c)=(2,5,1)$ in the $(-1,0,1)$ sector, where $\Omega'(2,5,1)=0$ and $\Omega''(2,5,1)=23544$. Note an integer factorization of 23544 is
\begin{equation}
23544=3\times 24\times 327
\end{equation}
where $3=b-a$, $24$ is the coefficient of $x^0$ in the Taylor expansion of $\Delta(x)^{-1}$ and $327$ is the coefficient of $w^1$ in the $(-1,1)$ sector of $\Phi_{10}^{-1}(y,z;w)$.

For $(a,b,c)=(3,4,1)$ in the $(-1,1,1)$ sector, we have 
\begin{equation}
\Omega''(3,4,1)-\Omega'(3,4,1)=105948=1\times 324\times 327
\end{equation}
where $1=b-a$, $324$ is the coefficient of $x^1$ in the Taylor expansion of $\Delta(x)^{-1}$. Both of these 2 examples show the following pattern(schematically) of wall-crossing
\begin{equation}\label{wc}
\Omega''-\Omega'=(b-a)\Omega_{1+3}\Omega_2
\end{equation}
where $\Omega_{1+3}$ means the number of states in the (1,3) subsystem and $\Omega_2$ is the appropriate expansion coefficient in (\ref{oneoverDelta}). Thus equation (\ref{wc}) exactly matches the structure of the 2-center wall crossing formula (\ref{Omfacttwo}), with $\Omega_{1+3}$ the correct degeneracy for the charge $\Gamma_1+\Gamma_2$ on $K3 \times T^2$!

The following two tables are a short summary of more numerical results in the $(r,s,t)=(-1,1,1)$ sector.

\begin{center}
\captionof{table}{$(-1,1,1)$ sector wall-crossing from $(b-a)\Omega_2\Omega_{1+3}$}
\begin{tabular}{|r|ccccc|} 
\hline 
&$\Omega''-\Omega'$& $b-a $ & $\Omega_2$ &$\Omega_{1+3}$& $(b-a)\times \Omega_2 \times \Omega_{1+3}$\\
\hline
(2,4,1)&211896 & 2 & 324& 327&211896 \\
(3,4,1)&105948 & 1 & 324& 327&105948\\
(2,5,1)&317844 & 1 & 324& 327&317844\\
(3,5,1) &211896 & 2 & 324& 327&211896\\
\hline
\end{tabular}
\end{center}

\begin{center}
\captionof{table}{$(-1,1,1)$ sector wall-crossing from $(c-a)\Omega_1\Omega_{2+3}$}
\begin{tabular}{|r|ccccc|} 
\hline 
&$\Omega''-\Omega$& $c-a $ & $\Omega_1$ &$\Omega_{2+3}$& $(b-a)\times \Omega_2 \times \Omega_{2+3}$\\
\hline
(2,1,4)&260658 & 2 & 1& 130329&260658 \\
(2,2,4)&418608 & 2 & 1& 209304&418608\\
\hline
\end{tabular}
\end{center}
where $130329$ and $209304$ are the coefficients of $yzv$ and $yzv^2$ in the expansion of $\Phi_{10}^{-1}$.

\section{Discussion} \label{sec:discussion}

In this paper we have put forward a conjecture for a precise counting function governing the three-center BPS solutions in type II string compactification on $K3 \times T^2$.  Support for our conjecture comes from correct behavior under
wall-crossing, and from the appearance of the known counting functions governing single and two-center solutions
($1/\Delta$ and $1/\Phi_{10}$) in appropriate degenerate limits.

\medskip
\noindent The paper raises a number of questions: 

\medskip
\noindent
$\bullet$ The objects we have described depend on a partition of the total charge $\Gamma$ into three 1/2-BPS charges $\Gamma_1,\Gamma_2,\Gamma_3$. The invariants $(r,s,t;a,b,c)$ defined in (\ref{rstabcdef})  depend on this partition. Thus the coefficients $\Omega(r,s,t;a,b,c)$ of $1/\chi_9$ do not count all BPS states with a given total charge $\Gamma$, but rather count a partition-dependent subset thereof. Our results suggest this subset is captured by a  microscopic model characterized by $(r,s,t;a,b,c)$, akin to the simple cyclic 3-node quiver quantum mechanics models studied in \cite{Denef:2007vg,Bena:2012hf,Lee:2012sc,Manschot:2013sya}. More specifically this should be a supersymmetric quantum mechanics model describing the D-brane systems of section \ref{sec:Dbranesetup}, in the spirit of for example the explicit models of \cite{Chowdhury:2014yca} describing D-branes on $T^6$. What is the precise microscopic model appropriate for our setup? More generally, one could ask if there exists a model-independent way of characterizing $\Omega(\Gamma_1,\Gamma_2,\Gamma_3)$. A natural physical object depending on charge partitions is the S-matrix; perhaps this may provide such a characterization along the lines of \cite{Harvey:1996gc}.

\medskip
\noindent
$\bullet$ We have loosely interpreted the coefficients of $1/\chi_9$ as BPS degeneracies, but did not provide a  definition in terms of a protected index. The standard index counting 1/4-BPS states in $\CN=4$ theories at generic points in the moduli space is the helicity supertrace $B_6$. However this cannot be the appropriate index counting the states of interest to us: with the exception of two-center bound states of 1/2-BPS black holes, multi-center bound states are BPS only on positive-codimension subspaces of the moduli space, and have too many fermionic zero
modes to contribute to $B_6$ \cite{Sen:2008ht,Dabholkar:2009dq,Sen:2009md,Ashoke,Kachru:2017yda}. Does there exist an index interpretation of the coefficients $\Omega(r,s,t;a,b,c)$ on suitable subspaces of the moduli space, perhaps along the lines of \cite{Sen:2009md}?

\medskip
\noindent
$\bullet$ Our analysis of wall crossing and its relation to contour choices did not reach the level of precision and generality of the prescriptions in \cite{Sen:2007vb,Sen:2007pg,Cheng:2007ch,Banerjee:2008yu} for extracting BPS degeneracies from $1/\Phi_{10}$ at a given point in the $T^2 \times K3$ moduli space. What is the analogous prescription for extracting moduli-dependent degeneracies from $1/\chi_9$?

\medskip
\noindent
$\bullet$ Is there a natural geometric way of understanding the origin of the genus-three Riemann surface associated with $1/\chi_9$? A geometric origin of the genus-two Riemann surface associated with $1/\Phi_{10}$ was suggested in \cite{Gaiotto} and further clarified in \cite{Banerjee:2008yu}. Higher genus generalizations of this construction have appeared in counting higher-torsion
dyons in ${\cal N}=4$ string theory \cite{Atishtwo,Atishthree} and BPS states  in geometrically engineered quantum field theories \cite{Hollowood:2003cv,Braden:2003gv}. These higher-genus Riemann surfaces are non-generic, however, as they are holomorphically embedded in $T^4$, and thus correspond to a three-dimensional subspace of the higher-genus Siegel upper-half space.
Understanding the relationship of our results with these constructions should be instructive. 

\medskip
\noindent
$\bullet$ The appearance of a degree three Siegel form counting three-center bound states  suggests
that there should be a higher genus generalization, with a degree four Siegel form counting four-center bound states
and so forth. Indeed the number of duality invariants of a $g$-center configuration equals $g+{g \choose 2} = \frac{1}{2} g(g+1)$, which equals the dimension of the genus-$g$ Siegel upper-half space.\footnote{Note that this is strictly larger than the dimension $3g-3$ of the complex structure moduli space of genus-$g$ Riemann surfaces when $g > 3$, so it is important that the form extends over the full Siegel upper-half space.}
At genus four we have a precise candidate, involving the Schottky form $J_8$ \cite{Morozov}.  Can
one make a uniform story capturing the physics at all genera?  A natural conjecture is that it involves the chiral genus $g$ bosonic string
partition function. 


\section*{Acknowledgements}

We thank M.\ Douglas, C.\ Vafa and M.\ Zimet for helpful and stimulating discussions.
S.K.\ and A.T.\ are grateful to the Aspen Center for Physics for hospitality while this work was in progress.
The research of 
F.D.\ and Z.S.\ is supported in part by the Department of Energy under contract DOE DE-SC0011941.
The research of  S.K.\ is supported in part by a Simons Investigator Award and by the National Science Foundation under grant number PHY-1720397.  The research of A.T.\ is supported by the National Science Foundation under
NSF MSPRF grant number 1705008.



\begin{thebibliography}{99}


\bibitem{StromingerVafa}
A.~Strominger and C.~Vafa,
 ``Microscopic origin of the Bekenstein-Hawking entropy,''
 Phys.\ Lett.\ B {\bf 379}, 99 (1996)
 doi:10.1016/0370-2693(96)00345-0
 [hep-th/9601029].

\bibitem{DVV}
 R.~Dijkgraaf, E.~P.~Verlinde and H.~L.~Verlinde,
 ``Counting dyons in N=4 string theory,''
 Nucl.\ Phys.\ B {\bf 484}, 543 (1997)
 doi:10.1016/S0550-3213(96)00640-2
 [hep-th/9607026].


\bibitem{Gaiotto}
 D.~Gaiotto,
 ``Re-recounting dyons in N=4 string theory,''
 hep-th/0506249.

\bibitem{Sen}
 A.~Sen,
 ``Black Hole Entropy Function, Attractors and Precision Counting of Microstates,''
 Gen.\ Rel.\ Grav.\  {\bf 40}, 2249 (2008)
 doi:10.1007/s10714-008-0626-4
 [arXiv:0708.1270 [hep-th]].


\bibitem{DMZ}
 A.~Dabholkar, S.~Murthy and D.~Zagier,
 ``Quantum Black Holes, Wall Crossing, and Mock Modular Forms,''
 arXiv:1208.4074 [hep-th].


\bibitem{Denef:2000nb} 
 F.~Denef,
 ``Supergravity flows and D-brane stability,''
 JHEP {\bf 0008}, 050 (2000)
 doi:10.1088/1126-6708/2000/08/050
 [hep-th/0005049].

\bibitem{Denef:2002ru}
F. Denef,
``Quantum quivers and Hall / hole halos,"
JHEP {\bf 0210}, 023 (2002)
[hep-th/0206072].


\bibitem{Bates:2003vx} 
 B.~Bates and F.~Denef,
 ``Exact solutions for supersymmetric stationary black hole composites,''
 JHEP {\bf 1111}, 127 (2011)
 doi:10.1007/JHEP11(2011)127
 [hep-th/0304094].

\bibitem{FKS}
 S.~Ferrara, R.~Kallosh and A.~Strominger,
 ``N=2 extremal black holes,''
 Phys.\ Rev.\ D {\bf 52}, R5412 (1995)
 doi:10.1103/PhysRevD.52.R5412
 [hep-th/9508072].



\bibitem{mooreattractor}
 G.~W.~Moore,
 ``Arithmetic and attractors,''
 hep-th/9807087.


\bibitem{Sen:2007vb} 
  A.~Sen,
  ``Walls of Marginal Stability and Dyon Spectrum in N=4 Supersymmetric String Theories,''
  JHEP {\bf 0705}, 039 (2007)
  doi:10.1088/1126-6708/2007/05/039
  [hep-th/0702141].

\bibitem{Dabholkar:2007vk} 
  A.~Dabholkar, D.~Gaiotto and S.~Nampuri,
  ``Comments on the spectrum of CHL dyons,''
  JHEP {\bf 0801}, 023 (2008)
  doi:10.1088/1126-6708/2008/01/023
  [hep-th/0702150 [HEP-TH]].

\bibitem{Sen:2007pg} 
  A.~Sen,
  ``Two centered black holes and N=4 dyon spectrum,''
  JHEP {\bf 0709}, 045 (2007)
  doi:10.1088/1126-6708/2007/09/045
  [arXiv:0705.3874 [hep-th]].

\bibitem{Cheng:2007ch} 
  M.~C.~N.~Cheng and E.~Verlinde,
  ``Dying Dyons Don't Count,''
  JHEP {\bf 0709}, 070 (2007)
  doi:10.1088/1126-6708/2007/09/070
  [arXiv:0706.2363 [hep-th]].

\bibitem{Banerjee:2008yu} 
  S.~Banerjee, A.~Sen and Y.~K.~Srivastava,
  ``Genus Two Surface and Quarter BPS Dyons: The Contour Prescription,''
  JHEP {\bf 0903}, 151 (2009)
  doi:10.1088/1126-6708/2009/03/151
  [arXiv:0808.1746 [hep-th]].
  
\bibitem{Sen:2008ht} 
A.~Sen,
  ``Wall Crossing Formula for N=4 Dyons: A Macroscopic Derivation,''
  JHEP {\bf 0807}, 078 (2008)
  doi:10.1088/1126-6708/2008/07/078
  [arXiv:0803.3857 [hep-th]].  
  
\bibitem{Dabholkar:2009dq} 
  A.~Dabholkar, M.~Guica, S.~Murthy and S.~Nampuri,
  ``No entropy enigmas for N=4 dyons,''
  JHEP {\bf 1006}, 007 (2010)
  doi:10.1007/JHEP06(2010)007
  [arXiv:0903.2481 [hep-th]].


\bibitem{Denef:2007vg} 
 F.~Denef and G.~W.~Moore,
 ``Split states, entropy enigmas, holes and halos,''
 JHEP {\bf 1111}, 129 (2011)
 doi:10.1007/JHEP11(2011)129
 [hep-th/0702146].

\bibitem{Bena:2012hf} 
 I.~Bena, M.~Berkooz, J.~de Boer, S.~El-Showk and D.~Van den Bleeken,
 ``Scaling BPS Solutions and pure-Higgs States,''
 JHEP {\bf 1211}, 171 (2012)
 doi:10.1007/JHEP11(2012)171
 [arXiv:1205.5023 [hep-th]].

\bibitem{Lee:2012sc} 
  S.~J.~Lee, Z.~L.~Wang and P.~Yi,
  ``Quiver Invariants from Intrinsic Higgs States,''
  JHEP {\bf 1207}, 169 (2012)
  doi:10.1007/JHEP07(2012)169
  [arXiv:1205.6511 [hep-th]].

\bibitem{Manschot:2013sya} 
 J.~Manschot, B.~Pioline and A.~Sen,
 ``On the Coulomb and Higgs branch formulae for multi-center black holes and quiver invariants,''
 JHEP {\bf 1305}, 166 (2013)
 doi:10.1007/JHEP05(2013)166
 [arXiv:1302.5498 [hep-th]].

\bibitem{BKMP}
A.~A.~Belavin, V.~Knizhnik, A.~Morozov and A.~Perelomov,
 ``Two and Three Loop Amplitudes in the Bosonic String Theory,''
 JETP Lett.\  {\bf 43}, 411 (1986)
 [Phys.\ Lett.\ B {\bf 177}, 324 (1986)].
 doi:10.1016/0370-2693(86)90761-6

\bibitem{Beilinson}
A.~A.~Beilinson and Y.~I.~Manin,
 ``The Mumford Form and the Polyakov Measure in String Theory,''
 Commun.\ Math.\ Phys.\  {\bf 107}, 359 (1986).


\bibitem{Atishthree}
 A.~Dabholkar, J.~Gomes and S.~Murthy,
 ``Counting all dyons in N =4 string theory,''
 JHEP {\bf 1105}, 059 (2011)
 doi:10.1007/JHEP05(2011)059
 [arXiv:0803.2692 [hep-th]].

\bibitem{Brunner:1999jq} 
  I.~Brunner, M.~R.~Douglas, A.~E.~Lawrence and C.~Romelsberger,
  ``D-branes on the quintic,''
  JHEP {\bf 0008}, 015 (2000)
  doi:10.1088/1126-6708/2000/08/015
  [hep-th/9906200].


\bibitem{Kachru:1999vj} 
 S.~Kachru and J.~McGreevy,
 ``Supersymmetric three cycles and supersymmetry breaking,''
 Phys.\ Rev.\ D {\bf 61}, 026001 (2000)
 doi:10.1103/PhysRevD.61.026001
 [hep-th/9908135].


\bibitem{Douglas:2000ah} 
  M.~R.~Douglas, B.~Fiol and C.~Romelsberger,
  ``Stability and BPS branes,''
  JHEP {\bf 0509}, 006 (2005)
  doi:10.1088/1126-6708/2005/09/006
  [hep-th/0002037].

\bibitem{Cremades:2003qj} 
 D.~Cremades, L.~E.~Ibanez and F.~Marchesano,
 ``Yukawa couplings in intersecting D-brane models,''
 JHEP {\bf 0307}, 038 (2003)
 doi:10.1088/1126-6708/2003/07/038
 [hep-th/0302105].

\bibitem{Chowdhury:2014yca} 
  A.~Chowdhury, R.~S.~Garavuso, S.~Mondal and A.~Sen,
  ``BPS State Counting in N=8 Supersymmetric String Theory for Pure D-brane Configurations,''
  JHEP {\bf 1410}, 186 (2014)
  doi:10.1007/JHEP10(2014)186
  [arXiv:1405.0412 [hep-th]].  


\bibitem{Cvetic:1995bj} 
 M.~Cvetic and A.~A.~Tseytlin,
 ``Solitonic strings and BPS saturated dyonic black holes,''
 Phys.\ Rev.\ D {\bf 53}, 5619 (1996)
 Erratum: [Phys.\ Rev.\ D {\bf 55}, 3907 (1997)]
 doi:10.1103/PhysRevD.53.5619, 10.1103/PhysRevD.55.3907
 [hep-th/9512031].



\bibitem{Bena:2006kb} 
 I.~Bena, C.~W.~Wang and N.~P.~Warner,
 ``Mergers and typical black hole microstates,''
 JHEP {\bf 0611}, 042 (2006)
 doi:10.1088/1126-6708/2006/11/042
 [hep-th/0608217].

\bibitem{deBoer:2008zn} 
 J.~de Boer, S.~El-Showk, I.~Messamah and D.~Van den Bleeken,
 ``Quantizing N=2 Multicenter Solutions,''
 JHEP {\bf 0905}, 002 (2009)
 doi:10.1088/1126-6708/2009/05/002
 [arXiv:0807.4556 [hep-th]].

\bibitem{Manschot:2010qz} 
 J.~Manschot, B.~Pioline and A.~Sen,
 ``Wall Crossing from Boltzmann Black Hole Halos,''
 JHEP {\bf 1107}, 059 (2011)
 doi:10.1007/JHEP07(2011)059
 [arXiv:1011.1258 [hep-th]].
  

\bibitem{Manschot:2014fua} 
 J.~Manschot, B.~Pioline and A.~Sen,
 ``The Coulomb Branch Formula for Quiver Moduli Spaces,''
 arXiv:1404.7154 [hep-th].  

\bibitem{Zimo}
Z. Sun, to appear.

\bibitem{Harvey:1996gc} 
  J.~A.~Harvey and G.~W.~Moore,
  ``On the algebras of BPS states,''
  Commun.\ Math.\ Phys.\  {\bf 197}, 489 (1998)
  doi:10.1007/s002200050461
  [hep-th/9609017].


\bibitem{Sen:2009md} 
  A.~Sen,
  ``A Twist in the Dyon Partition Function,''
  JHEP {\bf 1005}, 028 (2010)
  doi:10.1007/JHEP05(2010)028
  [arXiv:0911.1563 [hep-th]].

\bibitem{Ashoke}
 A.~Sen,
 ``BPS Spectrum, Indices and Wall Crossing in N=4 Supersymmetric Yang-Mills Theories,''
 JHEP {\bf 1206}, 164 (2012)
 doi:10.1007/JHEP06(2012)164
 [arXiv:1203.4889 [hep-th]].

\bibitem{Kachru:2017yda} 
  S.~Kachru and A.~Tripathy,
  ``BPS jumping loci and special cycles,''
  arXiv:1703.00455 [hep-th].


\bibitem{Atishtwo}
 A.~Dabholkar, D.~Gaiotto and S.~Nampuri,
 ``Comments on the spectrum of CHL dyons,''
 JHEP {\bf 0801}, 023 (2008)
 doi:10.1088/1126-6708/2008/01/023
 [hep-th/0702150 [HEP-TH]].






  


\bibitem{Hollowood:2003cv} 
  T.~J.~Hollowood, A.~Iqbal and C.~Vafa,
  ``Matrix models, geometric engineering and elliptic genera,''
  JHEP {\bf 0803}, 069 (2008)
  doi:10.1088/1126-6708/2008/03/069
  [hep-th/0310272].

\bibitem{Braden:2003gv} 
  H.~W.~Braden and T.~J.~Hollowood,
  ``The Curve of compactified 6-D gauge theories and integrable systems,''
  JHEP {\bf 0312}, 023 (2003)
  doi:10.1088/1126-6708/2003/12/023
  [hep-th/0311024].
  
  
\bibitem{Morozov}
A. Morozov, ``Explicit formulae for one, two, three and four loop string amplitudes,"
Phys. Lett. {\bf B184} (1987) 171.



  


\end{thebibliography}
\end{document}